\magnification=\magstep1
\hfuzz=6pt
\baselineskip= 16pt

$ $

\vskip 1in

\centerline{\bf Quantum Illumination}

\bigskip

\centerline{\it Seth Lloyd}

\centerline{W.M. Keck Center for Extreme Quantum Information Processing (xQIT)}

\centerline{MIT 3-160, Cambridge, MA 02139 USA}

\vskip 1cm

\noindent{\it Abstract:} The use of entangled light to illuminate
objects is shown to provide significant enhancements over unentangled
light for detecting and imaging those objects in the presence of 
high levels of noise and loss.  Each signal sent out
is entangled with an ancilla, which is retained.  Detection
takes place via an entangling measurement on the returning signal together
with the ancilla.  Quantum illumination with $e$ bits
of entanglement increases the effective signal-to-noise
ratio of detection and imaging by a factor of $2^e$, an 
exponential improvement over unentangled illumination.

\vskip 1cm
We are trying to detect the presence of an object by 
shining light in its direction and seeing if any is reflected back.  
Only a small percentage of
the light we shine is reflected.  The object
is immersed in noise and thermal radiation, 
so we have to distinguish whatever light is reflected from the 
noisy background.   The usual solution to this problem, as
in, e.g., radar or lidar, is to send a signal into the region where
the object may be, and to monitor the radiation returning from that
region to see if any trace of the signal can be detected
above the background noise.  Typically, the great majority of 
the signal is lost and only a tiny fraction returns, if any.

In the case of quantum bits, it is known that the sensitivity of 
detection processes can be enhanced by entangling the signal with an
ancilla, and by making an entangling measurement on the
returning radiation together with that ancilla [1].  Here,
we ask whether entanglement gives an enhancement
of sensitivity for quantum optical processes, and 
investigate the amount of such enhancement, if any.
The intuition
is that if we send out a signal photon that is entangled with
an ancillary photon, then when that photon comes back, if ever,
it will prove easier to recognize as the same photon that was
sent out.  This intuition will turn out to be correct,
{\it even though noise and loss completely destroy the entanglement
betwen signal and ancilla.}  Moreover, the entanglement-induced
enhancement of sensitivity is significant: it grows exponentially
with the number of bits of entanglement between signal and ancilla.
The exponential enhancement in effective signal-to-noise ratio
persists when we process returning light to construct an image of
the object.

In this paper, we present the simplest possible treatment of
quantum illumination, where signal and ancilla consist of
individual photons created, e.g., via spontaneous parametric
downconversion.  A more complete treatment of quantum illumination
in terms of generalized Gaussian states will be presented elsewhere. 
Suppose that we send a single photon at the object to be
detected.  If the object is not there, the signal that
we receive back consists of thermal and background noise. 
Even if the object is there, the great majority of 
time the photon is lost, and we receive only noise.
Once in a great while, the photon is reflected to us
in a perturbed form.  The dynamics corresponding
to this situation can be modelled as beam splitter with
reflectivity $\eta$, that mixes in the vacuum with the
signal state, followed by thermalization which mixes in
noise with average photon number $b$ per
mode.  No object corresponds to a beam
splitter with $\eta = 0$.  The presence of an object
corresponds to a beam splitter with a very small 
non-zero reflectivity.

We repeatedly send single photons at
the object, and try to detect reflected photons.  
The number of noise photons per mode will be taken to be small.  
This scenario corresponds, e.g., to single photons in the
optical regime directed at a distant target which is
bathed in thermal radiation at temperature significantly
below optical energies.  In this case,    
$b = (1 -e^{-\hbar\omega/kT})
e^{-\hbar\omega/kT}$ is the average number of thermal photons
per mode.  Our detector can distinguish between
$d$ modes per detection event: $d = W T$, where $W$
is the bandwidth of the detector and $ T$ is the 
length of the detection window. We take the time window
for detection sufficiently small that at most 
one noise photon is detected at a time, i.e., $db << 1$.  
Additional, non-thermal noise can be tolerated as well,
as long as fewer than one photon arrives per detection event.

First, consider the case of unentangled light.  We send
a single photon in the state $\rho$ toward the region where
the object might be.  The two different dynamics corresponding
to object there and object not there are as follows [2]:

\bigskip\noindent{\it Case (0),} object not there:
$ \rho  \rightarrow \rho_b\otimes \ldots \rho_b$, where
$\rho_b$ is the thermal state with $b$ photons per mode.
Because the average number of photons $bd$ received per
detection event is much less than 1, we can approximate
the thermal state as
$$  \rho_0   = (1-db) |vac\rangle \langle vac|
+ b\sum_{k=1}^d |k\rangle \langle k|.\eqno(1.0)$$
Here, $|vac\rangle$ is the vacuum state of the modes,
and $|k\rangle$ is the state where there is a single photon
in mode $k$ and no photons in the other modes.
Since $\| \rho_0  - \rho_b\otimes \ldots \rho_b \|_1
=(db)^2 + O((db)^3)$, we can safely replace the
exact thermal state by $\rho_0$ as long as we are
evaluating expression to lowest order in $db$.

\bigskip\noindent{\it Case (1),} object there:
$$\rho  \rightarrow  (1-\eta)  \rho_{0} + 
\eta \rho_{th},$$ 
where $\rho_{th}$ is the thermalized version of $\rho$.
It is straightforward to verify that $\|\rho - \rho_{th}\| =
db+O((db)^2)$.  Accordingly, if we are not interested
in terms of $O(\eta db)$, then we can safely approximate
$(1-\eta)  \rho_{0} + \eta \rho_{th}$ as
$$\rho \rightarrow \rho_1 = (1-\eta)  \rho_0 +
\eta \rho. \eqno(1.1)$$

Now we repeatedly send single photons in the state $|\psi\rangle$
to detect the presence or absence of the object. 
We look at the signal coming back to try to determine whether
the object is there or not.   The single shot minimum
probability of error is obtained by projecting
onto the positive part of $\rho_1 -\rho_0$ (see reference
[2] for a treatment of discrimination between different
dynamics in the case of qubits).  This
measurement consists simply of verifying whether
a returning photon is the state $|\psi\rangle$ or not.
If the measurement yields a positive result, then we
guess that the object is there.  If the result is
negative, then we guess that the object is not there.
The conditional probabilities of the outcomes `yes' and
`no' given the presence or absence of the object are
$$\eqalign{ p({\rm no} | {\rm not~there}) &= 1-b 
\quad p({\rm no} | {\rm there}) = (1-b)(1-\eta)\cr
p({\rm yes} | {\rm not~there}) &= b \quad \quad
p({\rm yes} | {\rm there}) =  b(1-\eta) + \eta.\cr} \eqno(2)$$ 

The number of trials required to reveal the presence or
absence of the object depends on the ratio $\eta/b$.
If $\eta/b > 1$, then a received photon is more likely
to be a signal photon than a noise photon: the signal to noise ratio is
greater than 1.  Call this regime, the good regime.
Similarly, if $\eta/b < 1$, then a received photon is more likely
to be a noise photon than a signal photon: the signal to noise ratio is
less than 1.  Call this regime, the bad regime. 
In the good regime, the probability that no photons have
been received after $n$ trials, given that the object
is not there, goes as $(1-b)^n$.  the probability that no photons have
been received after $n$ trials, given that the object
is there, goes as $(1-\eta)^n$.  That is,
the number of trials required to
detect the object, if there, goes as $1/\eta$: one
simply sends photons until one receives one back.
If the object is there, then one receives a photon
back considerably before one would expect a photon
given thermal photons alone.

In the bad regime, most of the photons received are
noise photons.  Here, one must count photons until
one can separate the thermal distribution with $b$
photons on average, from the distribution when the
object is there, which has $b+\eta$ photons on 
average.  Using the usual formulae for sampling the
outcomes of Bernoulli trials, one finds that
takes on the order of $8b/\eta^2$ photons on average to
distinguish between the presence or absence of the object
in the bad regime.

The optimal asymptotic minimum probability of error as the number of
trials $n$ gets large, is given
by the quantum Chernoff bound [3]: $ p_n(error) \approx   (1/2)Q^n,$ where 
$$Q= {\rm min}_s~ {\rm tr} \rho_0^{1-s} \rho_1^s. \eqno(3)$$
To lowest order in $\eta,b$, we have
$${\rho_0}^{1-s} = (1-db)^{1-s} |vac\rangle \langle vac | + b^{1-s} I,
\eqno(4.0)$$
where $I = \sum_k |k\rangle\langle k|$ is the identity operator
on the single photon subspace.  Similarly,
$${\rho_1}^s = (1-bd -\eta)^{s} |vac\rangle \langle vac | + 
b^s (I - |\psi\rangle\langle \psi|) +  (b+\eta)^s 
|\psi\rangle \langle \psi |. \eqno(4.1).$$ 
The trace of  $\rho_0^{1-s} \rho_1^s$ can be readily evaluated,
and the quantum Chernoff bound takes the form 
$$Q = {\rm min}_s~ \bigg(1-\eta s  + 
b\big(-1 + (1+{\eta\over b})^s \big) \bigg)  + O(b^2, \eta b).\eqno(5)$$  

Just as in the iterated single shot case,
equation (5) implies the existence of two regimes for 
evaluating the quantum Chernoff bound, depending on whether
$\eta > b$ (good) or $\eta < b$ (bad).  
The exact value of $s$ that minimizes $Q$ in
the good regime depends on the ratio $b/\eta$.  In the limit that 
$b/\eta << 1$, the minimum occurs for $s\rightarrow 1$, and  
$$Q =  1 -\eta +O((db)^2) \eqno(6)$$   
Comparing the quantum Chernoff bound with the minimum error
single shot bound above, we see that they yield the same
asymptotic error probability.  Accordingly, as before,
the optimal measurement needed to attain the quantum Chernoff
bound is simply to count the number of signal photons 
in the state $|\psi\rangle$
that return.  In the good regime, if the object is there, such a photon will
come back after $1/\eta$ trials on average, sooner than 
the $1/b$ trials expected if the object is not 
there.  

When $\eta < b$, a photon that comes
back is more likely to be a noise photon than a signal photon.
As before, the $\eta < b$ regime is
the `bad' regime where the signal to noise
ratio less than one.  Evaluating equation (5) in this regime, we find 
$$Q =    1 - {\eta^2 \over 8 b } + O({\rm max}~ b^2, \eta b). \eqno(7)$$
(Note that $\eta^2/b >> \eta b, b^2$, because $b << 1$.)
Once again, the quantum Chernoff bound coincides asymptotically
with the single shot minimum error rate: 
the optimal measurement is simply to count returning
photons in the state $|\psi\rangle$
to see if the number of photons returned is significantly different
from the number expected if we were receiving only thermal
radiation.  Here, because of the low signal to noise ratio,
the signal gives only a small shift
in the average number of received photons, and
$ \approx 8b/\eta^2$ photons must be sent to
detect the object with high probability.  

Note that the error probabilities for detecting
the presence or absence of the object do not depend on the 
number of signal and detector modes $d$.  The number
of modes doesn't matter because all modes
other than the one in which the photon is sent are in a
thermal state.  These other modes give us no information about the presence
or absence of the object.  To detect the object,
we need only monitor the mode in which we
sent the photon to see if more than the expected number
of photons come back.

\bigskip\noindent{\it Sending entangled photons}

Now let us look at the effect of entanglement on our ability
to detect the object.  Construct the entangled state
$|\psi\rangle_{SA} = (1/\sqrt d) \sum_k |k\rangle_S |k\rangle_A$
for signal photon and ancilla photon, and send the signal photon to
towards where the object is likely to be.   This state
can be constructed, for example, by taking the output of 
a spontaneous parametric downconverter in the low-photon
number regime, matching its time-bandwidth product to the
time-bandwidth product of the detector,
and selecting out the one signal photon/ one
idler photon sector (as will be seen below, photodetection
at the receiver can automatically postselect this state).  
The signal photon is sent off
and the idler photon is retained as the ancilla.

The two different dynamics corresponding to object or no object
now take a slightly different form, as the
state of the ancilla must be included.  If the signal photon is lost,
the ancilla photon goes to the completely mixed state: 

\bigskip\noindent{\it Case (0),} object not there:
$$|\psi\rangle_{SA}\langle \psi|  
\rightarrow \rho_{SA0} =  \rho_{0} \otimes {I_A\over d} = 
\big( (1-db) |vac\rangle_
S\langle vac| + b I_S \big) \otimes {I_A\over d} + O(b^2).\eqno(8.0)$$

\bigskip\noindent{\it Case (1),} object there:
$$ |\psi\rangle_{SA}\langle \psi| 
\rightarrow \rho_{SA1} = (1-\eta) \rho_{SA0}  +
\eta |\psi\rangle_{SA}\langle \psi| + O(b^2, \eta b).\eqno(8.1)$$

\bigskip\noindent As before, the single-shot minimum 
error probability is obtained by projecting onto the
positive part of $(\rho_{SA1} - \rho_{SA0})$, which in
turn simply corresponds to determining whether any
returning photon is in the state $|\psi\rangle_{SA}$.

A detailed treatment of entangled photodetection lies
outside the scope of this paper.  Although such
entangling measurements are likely to prove difficult,
they are certainly allowed by the laws of physics.
Take the case where
$|\psi\rangle_{SA}$ is the single-photon pair
output of a spontaneous parametric downconverter, for example.
In this state, the signal and idler modes are anticorrelated in
momentum, so that $\omega_k^S + \omega_k^I = \omega$,
where $\omega$ is the pump frequency.  In addition, because of
the form of the entanglement, the signal and idler modes
are correlated in time of arrival, so that 
both photons will arrive at a receiver at the same time.

To check if the two photons are in the state
$|\psi\rangle_{SA}$, one needs to verify
both frequency anticorrelation and time of arrival
correlation between the photons.  More explicitly,
the entangling measurement on signal and ancilla
must verify that signal and ancilla frequencies
sum to $\omega$, without distinguishing between the
different modes $k$; then photodetection must be carried
out to check that both signal and ancilla arrive at
the detector at the same time.  For example, the detector
could consist of an ensemble of atoms which can only make two-photon
transitions with sum frequency $\omega$.  Because
of the positive correlation in time of arrival, the
absorption rate of the two entangled photons is linear
in the photon flux density rather than quadratic [4] --
the entangled state is much more likely to induce
such a transition than an unentangled state.  The
ensemble can then be queried using, e.g., a cycling
transition, to determine if any two-photon transition
has taken place.  Note that such two-photon detection
retroactively post-selects the single photon pair
state out of the spontaneous parametric amplifier output 
state in the low flux regime. 

Making the optimal single-shot measurement and
evaluating the conditional error probabilities
for the entangled case yields
$$\eqalign{ p_e({\rm no} | {\rm not~there}) &= 1-{b\over d}
\quad p_e({\rm no} | {\rm there}) = (1-{b\over d})(1-\eta)\cr
p_e({\rm yes} | {\rm not~there}) &= {b\over d} \quad \quad
p_e({\rm yes} | {\rm there}) = (1-\eta){b\over d} + \eta.\cr} \eqno(9)$$
It is immediately seen that the effect of entanglement
is to reduce the effective noise from $b$ to $b/d$.
This reduction reflects the fact that in the entangled case 
a noise photon together with the fully 
mixed ancilla is $d$ times less likely to be confused 
for a signal photon entangled with the ancilla,
than a noise photon is likely to be confused with
a signal photon in the unentangled case. 
Entanglement reduces the effective signal to noise
by a factor of $d$.

The single-shot minimum error probability for
the entangled case 
coincides with the asymptotic minimum error probability
by evaluating the quantum Chernoff bound, 
$Q = {\rm min}_s~ {\rm tr}~ \rho_{SA0}^{1-s}~ \rho_{SA1}^s.$
The roots of the density matrices can be evaluated and are only
slightly more complicated than before:
$$\eqalignno{
\rho_{SA0}^{1-s} &= \rho_S^{1-s}\otimes{ I_A \over d^{1-s}}\cr
&=\big( (1-db)^{1-s} |vac\rangle_S \langle vac | + b^{1-s} I_S \big)
\otimes { I_A \over d^{1-s}}.   &(10.0) \cr
\rho_{SA1}^s &= 
\big( (1-\eta) \rho_{SA0} + \eta  |\psi\rangle_{SA}\langle \psi| \big)^s\cr  
&=(1-bd -\eta)^{s} |vac\rangle_S \langle vac | \otimes{ I_A \over d^s}\cr
& +
(1-\eta)^s\bigg({b\over d} \bigg)^s 
\big(I_S\otimes I_A - |\psi\rangle_{SA}\langle \psi|\big) +	
\big((1-\eta){b\over d}+\eta\big)^s
|\psi\rangle_{SA} \langle \psi |. &(10.1)\cr} $$
Taking the trace of $\rho_{SA0}^{1-s}~ \rho_{SA1}^s$, we obtain
$$\eqalign{ Q&= {\rm min}_s~  
 (1-\eta)^s \bigg( 1 + {b\over d} 
\big( -1 + ( 1 + {\eta d \over (1-\eta)b})^s \big) \bigg) \cr 
&= {\rm min}_s~ 
\bigg(1 -\eta s + {b\over d} \big( -1 + (1+{\eta d\over b})^s \big)\bigg) 
+O(b^2, \eta b). \cr}
\eqno(11)$$
Equation (11) confirms that the effect of entanglement
is to reduce the effective noise from $b$ to $b/d$.

Comparing the quantum Chernoff bound for entangled
states, equation (9), the quantum Chernoff bound for
unentangled states, equation (4), we see that there are once more two regimes.
The good regime now occurs when $\eta d/b >1$.  In this regime, 
the quantum Chernoff bound is 
$$Q \approx 1- \eta \eqno(11)$$ 
in the limit $\eta d/b >> 1, s \rightarrow 1$.  Comparing
the entangled case to the unentangled case above, we see
that the quantum Chernoff bound is the {\it same} in
the good regime in both cases, but the
good regime extends $d$ times further in the entangled case
than in the unentangled case, where the good regime occurred
for $d/b > 1$.  

The extension of the good regime via the use of entangled
photons can be understood as follows.  
As before, the quantum Chernoff bound coincides with repeated
single-shot minimum error measurements, showing that the
the optimal detection
strategy is to measure any incoming photon together with
the ancilla to see if the two photons are in the state
$|\psi\rangle_{SA}$.   If the photon that returns is
the signal photon, then it will pass the test.  If the
photon that returns is a noise photon, then the ancilla
is in the state $I_A/d$,  and the noise photon together with
fully mixed ancilla are $d$ times less likely 
to be found in the state $|\psi\rangle_{SA}$.  A noise
photon in the entangled case is $d$ times less likely
to pass the test and be confused as a signal photon than
a noise photon in the unentangled
case.  In other words, the presence of entanglement makes it $d$
times harder for a noise photon to masquerade as a signal 
photon.  Entanglement effectively enhances the 
signal to noise ratio by a factor of $d$.

The bad regime for the entangled case occurs for $\eta d/b < 1$.
In this case the quantum Chernoff bound occurs for $s\approx 1/2$ and is 
$$Q = 1 - {\eta^2 d \over 8 b}  +O(b^2, \eta b). \eqno(13)$$
Comparing with equation (7) for the quantum Chernoff bound
in the unentangled bad regime, we see that the entangled bound
is $d$ times better than the unentangled bound: quantum 
illumination reduces the number of trials needed to detect
the object by a factor $d$.  Entanglement 
effectively enhances the signal to noise ratio by the degree
of entanglement, even in the bad regime.  

The fact that entanglement yields an enhancement in the bad regime
is particularly interesting, because in the bad regime the combination
of noise and loss insures that {\it
no entanglement between signal and ancilla survives}, an effect
that also appears in the qubit case [1].  Nonetheless,
even though signal and ancilla are unentangled at the detector,
a noise photon still finds it $d$ times harder to masquerade
as a signal photon entangled with an ancilla photon.

\bigskip\noindent{\it Imaging}

Entanglement effectively enhances the signal to noise ratio of
detection by a factor of $d$, where $d$ is the number of entangled
modes.  Measured in terms of $e =$ the number of e-bits of entanglement, 
the enhancement is 
$$d = 2^{e}. \eqno(14)$$
The enhancement is exponential in the number e-bits.

This enhancement persists in imaging.  Suppose that one 
images an object by illuminating it in a point-by-point 
fashion.  Light is focused on the object to the
minimum spot size allowed by the Rayleigh bound.
Light reflecting from that spot is focused on
an image plane using a lens.  The dimensions of
the spot's image are also determined by the Rayleigh bound.
Photodetection and postprocessing of the signal at the
image plane then allow the amount of light reflected
from that spot on the object to be measured.  The light is now focused
on another spot and the process is repeated until the entire
object has been imaged.  
Such a method can yield a classical enhancement
of $\sqrt n$ over the Rayleigh limit, where $n$ is the number
of photons collected for each point [5-6].  

A key limitation of this semiclassical imaging
method is the presence of noise in the form of
light originating from points other
than the point that is being imaged at that moment.
If the light used to image the object is entangled with an
ancilla in the fashion described above, and entangled detection of
signal and ancilla is performed at the image plane, then once again noise
photons find it $d$ times harder to masquerade as
signal photons.  The same quantum illumination
techniques that allow an exponential enhancement of
the signal to noise ratio for detection also provide
a potential exponential enhancement of imaging.  In
imaging, however, the Rayleigh limit on the
spatial extent of the modes at the image plane,
together with fact that the entangled
photo-detection is a two photon process raise
serious questions about the efficiency of photodetection,
despite the positive correlation in photon time of arrival [4].
Considerable further analysis of multiple quantized
spatial modes must be performed to
identify the scenarios in which quantum illumination
provides an actual practical advantage over ordinary
illumination for imaging.

\vfill\eject

\bigskip\noindent{\it Discussion}

Quantum illumination is a potentially powerful technique
for performing detection and imaging, in which signal
is entangled with an ancilla, and entangling measurements
are made at the detector.  Entanglement enhances the
effective signal to noise ratio because a noise photon has
a $d$ times harder time masquerading as an entangled
signal photon, compared with a noise photon masquerading
as an unentangled signal photon. The enhancement of
sensitivity and effective signal-to-noise ratio
that quantum illumination provides is exponential in
the number of bits of initial entanglement, and persists
even in the presence of large amounts of noise and loss,
when no entanglement survives at the receiver. 
Many practical questions remain, notably, 
how can the requisite entangled measurements be performed efficiently? 
Does the enhancement persist at higher noise temperatures
and for larger numbers of photons in the signal?   What
are the maximum enhancements obtainable via quantum
illumination over all possible input states, including Gaussian states? 
These questions and many others must be answered before
quantum illumination can prove itself useful in practice.

\vskip 1cm
\noindent{\it Acknowledgements:}   This work was supported by the
W.M. Keck foundation and by DARPA.  The author would like to thank
Baris Erkmen, Vittorio Giovannetti, Saikat Guha, Lorenzo Maccone,
Stefano Pirandola, Jeffrey Shapiro, Si-Hui Tan, Mankei Tsang, and
Horace Yuen for useful discussions.

\vskip 1in
\noindent{\it References:}

\bigskip\noindent[1]
M.F. Sacchi, {\it Phys. Rev. A} {\bf 71}, 062337 (2005);
arXiv:quant-ph/0505183.

\bigskip\noindent[2] W.H. Louisell, {\it Quantum Statistical Properties
of Radiation}, Wiley, 1990.

\bigskip\noindent[3] 
K.M.R. Audenaert, J. Calsamiglia, Ll. Masanes, R. Munoz-Tapia, 
A. Acin, E. Bagan, F. Verstraete,
{\it Phys. Rev. Lett.} {\bf 98}, 160501 (2007);
arXiv:quant-ph/0610027.

\bigskip\noindent[4] H.-B. Fei, B.M. Jost, S. Popescu, B.E.A. Saleh,
M.C. Teich, {\it Phys. Rev. Lett.} {\bf 78}, 1679 (1997).
 
\bigskip\noindent[5] 
S. J. Bentley and R. W. Boyd, {\it Opt. Express} {\bf 12}, 5735 (2004).

\bigskip\noindent[6]
A. Peier, B. Dayan, M. Vucelja, Y. Silberberg, and A. A. Friesem, {\it Opt. 
Express} {\bf 12}, 6600 (2004).

\vfill\eject\end